# Software Fault Localization Based on Multi-objective Feature Fusion and Deep Learning

Xiaolei Hu[a], Dongcheng Li[b,*], W. Eric Wong[c], Ya Zou[a]

[a] *School of Computer Science, China University of Geosciences, 430074, Wuhan, China*
[b] *Department of Computer Science, California State Polytechnic University - Humboldt, 95521, Arcata, USA*
[c] *Department of Computer Science, University of Texas at Dallas, 75080, Richardson, USA*

Software fault localization remains challenging due to limited feature diversity and low precision in traditional methods. This paper proposes a novel approach that integrates multi-objective optimization with deep learning models to improve both accuracy and efficiency in fault localization (FL). By framing feature selection as a multi-objective optimization problem (MOP), we extract and fuse three critical fault-related feature sets—spectrum-based, mutation-based, and text-based features—into a comprehensive feature fusion model. These features are then embedded within a deep learning architecture, comprising a multilayer perceptron (MLP) and gated recurrent network (GRN), which together enhance localization accuracy and generalizability. Experiments on the Defects4J benchmark dataset with 434 faults show that the proposed algorithm reduces processing time by 78.2% compared to single-objective methods. Additionally, our MLP and GRN models achieve a 94.2% improvement in localization accuracy compared to traditional FL methods, outperforming state-of-the-art deep learning-based FL method by 7.67%. Further validation using the PROMISE dataset demonstrates the model's generalizability, showing a 4.6% accuracy improvement in cross-project tests over state-of-the-art deep learning-based FL method.

## 1 INTRODUCTION

Currently, software fault localization faces two major challenges [1]: first, existing feature extraction techniques are inadequate in capturing software fault information, and second, many existing models suffer from inaccurate fault localization. In traditional software fault localization, reliance is typically placed on single-type features as guidance, leaving a large amount of feature information unexplored. Even though some advanced architectures emphasize extracting multiple types of features during the feature extraction process, they focus solely on dynamic fault localization features, those derived from information collected at runtime, while overlooking static textual features of the program for subsequent analysis. This paper considers both dynamic and static feature information in program faults and introduces a multi-objective optimization algorithm to address issues such as long algorithm runtime and computational redundancy that may arise from multidimensional fault feature calculations.

Deep learning, known for its adaptability and robustness in handling diverse data, has been increasingly applied to software fault localization. This approach has demonstrated improved accuracy over traditional fault localization techniques. However, it faces several challenges, including susceptibility to local minima and slower training speeds, which can result in the omission of critical error information. Furthermore, deep learning frameworks often struggle with time efficiency, particularly when applied to large-scale, real-world fault datasets, where processing times can increase significantly. To address those problems, this paper designs two deep learning fault localization models, combining the feature subsets extracted by the multi-objective fusion algorithm with improved multilayer perceptron (MLP) and gated recurrent unit (GRU), providing a new multi-dimensional feature processing approach for software fault localization.

The multi-objective feature fusion (MOFF) algorithm addresses the issues of static feature loss and feature information redundancy in software fault localization by transforming the feature selection problem into a multi-objective optimization problem (MOP). Simultaneously, the multi-objective feature fusion algorithm is integrated into deep learning models to enhance the accuracy and stability of fault localization models.

The contributions of this paper are as follows:

(1) To address the issue of missing feature information in software fault localization, this study selects three typical effective features: spectral fault features, mutation fault features, and textual fault features. This study transforms the problem into a multi-objective optimization issue, chooses effective features from multiple dimensions, and develops a multi-objective feature fusion algorithm using methods such as voting and weighting.

(2) To improve the accuracy of software fault localization, this study develops a fault localization model using deep learning techniques. Specifically, we design and implement frameworks based on multilayer perceptron (MLP) and recurrent neural network (RNN) models, incorporating a multi-objective feature fusion algorithm during the feature extraction phase.

The remaining sections of this paper are structured as follows: Section 2 reviews related research. Section 3 introduces the proposed fault localization approach based on multi-objective feature fusion. Section 4 proposes the fault localization models based on deep learning. Section 5 validates the performance of the proposed methods and models through experiments. Finally, the research work is summarized, and future directions are outlined in the concluding chapter.

## 2 RELATED STUDIES

With the increasing scale of modern software systems, traditional fault localization techniques are no longer effective in accurately identifying faults [2-3]. Consequently, in recent years, numerous advanced and effective fault localization techniques have emerged, such as those based on multi-objective optimization and deep learning [4-6]. These technologies optimize fault localization accuracy by combining multiple objective functions or leverage the powerful feature extraction capabilities of deep learning models to accurately identify fault sources from vast amounts of data.

### 2.1 Program Fault Localization Based on Multi-objective Optimization

Multi-objective optimization algorithms are a class of important optimization methods used to solve optimization problems involving multiple conflicting objectives [7-8]. The application of multi-objective optimization algorithms in software fault localization stems from the inherent trade-off between accuracy and recall. These algorithms achieve more robust fault localization by providing a set of Pareto optimal solutions to balance these two metrics.

The performance of software fault localization is influenced by various stages within the model, including data preprocessing, feature extraction, and model construction [9]. Among these, feature selection is a crucial step in data mining and machine learning, aiming to identify the most informative features from raw data to enhance model performance and mitigate the impact of the dimensionality curse. Commonly used feature selection methods include filter methods, wrapper methods, and embedded methods [10]. Embedded methods integrate feature selection into the training process of the learning algorithm, performing feature selection concurrently with model training, providing good efficiency and accuracy. These methods all have limitations to varying degrees. In recent years, many studies have proposed applying multi-objective optimization algorithms to the field of fault localization and feature extraction methods [11-13], achieving good results.

Wu et al. [14] examined the role of multi-objective optimization algorithms in feature selection for fault localization by using false alarm rates and localization probabilities as dual objectives in software testing. Gu et al. [15] emphasized the importance of selecting relevant features to improve model performance by expanding the particle search space in the particle swarm optimization algorithm to accommodate more feature subsets. Adline et al. [16] proposed a two-phase approach using the BMCBW algorithm for test suite minimization and an improved Long Short-Term Memory (LSTM) for fault validation, achieving better accuracy than existing models.

Multi-objective optimization algorithms can effectively reduce redundant features in software fault data, improving feature selection performance and the time efficiency of models, thereby saving computational resources. Compared to exhaustive or random search, heuristic search algorithms find the ideal feature subsets faster but may fall into local optima, requiring manual parameter tuning, which increases complexity and costs.

### 2.2 Program Fault Localization Based on Deep Learning

Program fault localization using deep learning has emerged as a prominent and actively researched topic, driven by advancements in adaptive algorithms that leverage data-based models for improved accuracy in identifying fault locations [17-19]. Wong and Qi [20] proposed a fault localization technique based on BP neural networks. In this fault localization model, the BP neural network learns the relationship between test case coverage and execution results and uses virtual test cases to predict the suspicious value of each statement to determine the fault location.

However, due to the problems of local minima in BP neural networks, Wong et al. [21] proposed an alternative method based on Radial Basis Function (RBF) networks, which are less affected by these problems and have a faster learning rate. In addition to BP neural networks and RBF network methods, Fu et al. [22] have proposed using weakly supervised deep learning for fault localization, such as the MetaFL method. This method successfully extends deep learning-based fault localization from supervised learning to weakly supervised learning and performs well in fault localization experiments. However, MetaFL was only tested on seven small-scale datasets and has not been validated on more widely adopted public test sets, limiting its generalizability.

To address the poor performance of neural networks under conditions of insufficient fault information, Li et al. [23] have proposed a deep learning fault localization framework, DeepRL4FL. The core idea is to treat fault localization as an image recognition problem, improving fault localization accuracy by enhancing the code coverage matrix and combining it with statement dependencies. However, there is still room for improvement in feature selection. As the feature dimensions continue to increase, traditional learning ranking algorithms struggle to automatically identify effective potential features. Li et al. [7] proposed a deep learning fault localization method, DeepFL, which can automatically learn effective existing and potential features to achieve accurate fault localization. DeepFL significantly outperforms the state-of-the-art TraPT in cross-project fault localization.

## 3 FAULT LOCALIZATION METHOD BASED ON MULTI-OBJECTIVE FEATURE FUSION

The overall process of the multi-objective feature fusion algorithm is illustrated in Fig. 1 and consists of the following four main components:

(1) Feature Extraction: Extract three key types of features from the faulty program, namely spectrum features, mutation features, and text features.

(2) Feature Selection: After extracting a large number of features, use multi-objective optimization algorithms to filter and select promising features.

(3) Feature Fusion: To integrate the advantages of different features and enhance the accuracy of fault diagnosis, feature fusion is performed using voting and weighting methods.

(4) Model Training and Evaluation: To assess the time efficiency (timeliness) and accuracy of the algorithm in fault localization, we establish two types of evaluation metrics: time efficiency and accuracy.

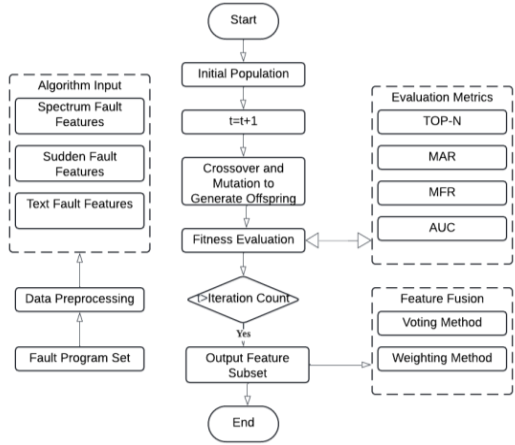

Fig. 1. Overall Process of the Multi-Objective Feature Fusion Algorithm

### 3.1 Encoding Design

In the feature selection stage of the software fault localization problem, three types of multi-objective optimization algorithms are used, all of which adopt a binary-coded chromosome encoding scheme. As shown in Fig. 2, assume there are five fault features: Feature A, Feature B, Feature C, Feature D, and Feature E. Each chromosome is represented with binary encoding, where each bit indicates whether a feature is selected (1) or discarded (0). For example, chromosome 10101 means that Features A, C, and E are selected, while Features B and D are discarded.

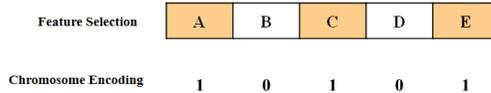

Fig. 2. Schematic Diagram of Fault Feature Chromosome Encoding

### 3.2 Operator Design

To better adapt to the binary classification characteristics of the fault localization problem, this paper employs uniform crossover, where the gene exchange probability is the same for each locus, thereby increasing the diversity of offspring. Let two chromosomes be $p_1$ and $p_2$, with binary encodings of $\{p_{11}, p_{12}, \ldots, p_{1n}\}$ and $\{p_{21}, p_{22}, \ldots, p_{2n}\}$, respectively. The two offspring generated by crossover are denoted as $c_1$ and $c_2$, with a gene exchange probability of $r$. The offspring produced by the crossover operation can be expressed as:

$$c_{1i} = \begin{cases} p_{1i}, & \text{if } r < 0.5 \\ p_{2i}, & \text{otherwise} \end{cases} \tag{1}$$

$$c_{2i} = \begin{cases} p_{2i}, & \text{if } r < 0.5 \\ p_{1i}, & \text{otherwise} \end{cases} \tag{2}$$

where $i$ is the gene position, and $r$ is a random number in the range [0,1]. The mutation operator further increases the diversity of the population by randomly altering some genes in the chromosome, preventing the algorithm from getting stuck in local optima. The mutation operator design considers the binary classification nature of the fault localization problem, specifically by setting a mutation probability that is related to fitness. Individuals with poorer fitness have a higher probability of mutation. The mutation probability can be expressed as:

$$P_m = \frac{f_{max} - f_i}{f_{max} - f_{min}} \tag{3}$$

where $f_i$ is the fitness value of an individual, and $f_{max}$ and $f_{min}$ represent the highest and lowest fitness values in the population, respectively. For binary encoding, the mutation operation is achieved by flipping a gene value at a certain locus (i.e., changing 0 to 1 or 1 to 0). To maintain population diversity and enhance the global search capability of the algorithm, the mutation operator design is expressed as:

$$c_i' = \begin{cases} 1 - c_i, & \text{if } r < P_m \\ c_i, & \text{otherwise} \end{cases} \tag{4}$$

where $c_i'$ is the new offspring chromosome generated by mutation, and $r$ is a random number in the range [0,1].

Crowding distance is used to measure the sparsity of individuals in the objective space and is an important criterion for selecting parents and evaluating individual diversity. The objective values of each individual are calculated for each objective function, and the individuals are sorted by these objective values. It is important to set the crowding distance of boundary individuals to infinity to ensure the preservation of boundary solutions. For each non-boundary individual, the crowding distance is calculated using the formula:

$$d_i = \sum_{j=1}^{M} \left( \frac{f_{j(i+1)} - f_{j(i-1)}}{f_j^{max} - f_j^{min}} \right) \tag{5}$$

where $M$ is the number of objective functions, $f_{j(i+1)}$ and $f_{j(i-1)}$ are the objective values of the individuals before and after individual $j$ in the sorted list for objective, $f_j^{max}$ and $f_j^{min}$ are the maximum and minimum values for objective $j$.

### 3.3 Fitness Evaluation

The accuracy of fault localization is usually measured by the ratio of the number of correctly localized faults to the total number of faults, with the accuracy calculation Acc process given by the formula:

$$Acc = \frac{N_{correct}}{N_{total}} \times 100\% \tag{6}$$

where $N_{correct}$ represents the number of true positives, $N_{total}$ the number of true negatives. Since the software fault localization problem studied in this paper is a classification problem, the confusion matrix for classification is shown in Table 1.

Table 1. Confusion Matrix for Fault Localization Problem

| Prediction vs. Actual | Fault (Prediction) | No Fault (Prediction) |
|---|---|---|
| Fault (Actual) | True Positive (TP) | False Positive (FP) |
| No Fault (Actual) | False Negative (FN) | True Negative (TN) |

Stability is usually evaluated by calculating the standard deviation of accuracy across multiple datasets or parameter settings:

$$Stability = \sqrt{\frac{1}{N} \sum_{i=1}^{N} (Acc_i - \overline{Acc})^2} \tag{7}$$

where $Acc_i$ is the accuracy under the $i$-th dataset or parameter setting, $\overline{Acc}$ is the average accuracy across all datasets or parameter settings, and $N$ is the number of datasets or parameter settings. A smaller stability indicator suggests that the algorithm performs more consistently under different conditions.

In addition to accuracy and stability, the runtime of the algorithm is also an important consideration. Typically, the runtime can be represented by the total time taken by the algorithm:

$$Time = T_{total} = T_{avg} \times N_{instance} \tag{8}$$

where $T_{avg}$ is the average processing time per program instance, and $N_{instance}$ is the total number of instances. By optimizing the execution efficiency of the algorithm, the maintenance cost of the fault localization system can be reduced.

### 3.4 Voting Method

To address the problem of an overly large solution space in multi-objective optimization algorithms, it is necessary to further filter out effective feature subsets. The Voting Method is a commonly used feature fusion technique [24], with its core idea being to determine the final feature subset by voting on different fault feature subsets. For example, consider the six fault feature subsets {f1, f2, f3, f4, f5, f6}. After three rounds of voting, the results might be {f1, f4, f6}, {f2, f4}, and {f2, f4, f6}. In this voting process, feature f1 appears once, feature f2 appears twice, feature f4 appears three times, and feature f6 appears twice. If we retain the top three feature subsets, the final feature subset would be {f2, f4, f6}.

### 3.5 Weighting Method

The feature subset selected by the Voting Method does not form an ordered sequence based on effectiveness. To further emphasize the most effective features within the subset, different weights are assigned to each feature. For instance, suppose the feature subset ultimately selected by the multi-objective optimization algorithm is {f2, f4, f6}, with corresponding weights of 0.5, 1, and 1.5. The fused feature subset would then be {0.5*f2, 1*f4, 1.5*f6}, where the numbers represent the weight coefficients of the features. The ordered feature subset, after sorting by weight coefficients, would be {1.5*f6, 1*f4, 0.5*f2}.

After considering both multi-objective feature fusion algorithms, this study opts to embed both methods into the algorithm for feature subset extraction to achieve complementary effects between the different feature result subsets.

## 4 DEEP LEARNING-BASED FAULT LOCALIZATION MODEL

The deep learning-based software fault localization model proposed in this study, which integrates multi-objective feature fusion and multi-objective optimization, is illustrated in Fig. 3. The model is divided into four parts: feature extraction, feature selection, fault localization, and performance evaluation, together forming the foundational framework and data flow of the entire software fault localization model.

### 4.1 Static Fault Localization Features

To address the issue of insufficient feature information in dynamic fault localization models, this study introduces text-based fault localization features as supplementary information during the feature extraction phase. The static fault localization features primarily include the number of lines of code, variable and operator counts, and branch path counts. Number of Lines of Code refers to the total number of lines in the code file. The calculation formula for the number of lines of code is as follows:

$$L = \sum_{i=1}^{n} L_i \tag{9}$$

where $L_i$ represents the length of the *i*-th line of code.

Variables refer to the objects operated upon, while operators are symbols or keywords used to perform operations. The calculation formula for variable and operator counts is as follows:

$$N_{op} = \sum_{i=1}^{n} N_{op}^i \tag{10}$$

$$N_{opr} = \sum_{i=1}^{n} N_{opr}^i \tag{11}$$

where $N_{op}^i$ represents the number of variables in the *i*-th line of code, and $N_{opr}^i$ represents the number of operators in the *i*-th line of code.

The Control Flow Graph (CFG) describes the control flow relationships between statements during program execution, i.e., the execution paths of the program. In CFG, nodes represent basic blocks and basic sequences of consecutive statements, while directed edges represent the control flow transitions between basic blocks.

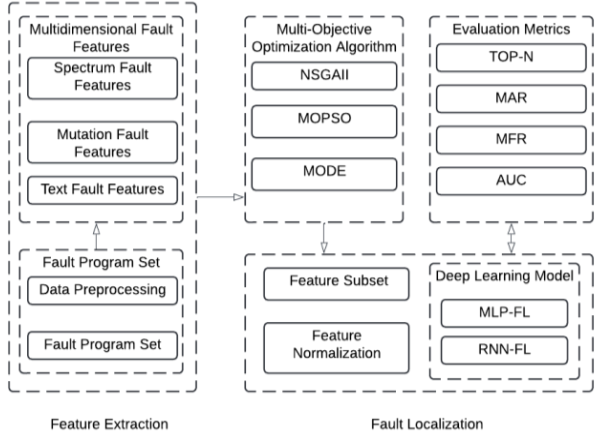

Fig. 3. Overall Process of the Deep Learning Fault Localization Model

Table 2. Static Feature Information for Statements in the Test Program

| Statement | Number of Branch Paths | Number of Variables | Number of Symbols |
|---|---|---|---|
| int x, y, z, m; | 3 | 4 | 4 |
| S1: input x, y, z; | 3 | 3 | 3 |
| S2: m = z; | 3 | 2 | 2 |
| S3: if (y < z) | 3 | 2 | 3 |
| S4: if (x<y) | 2 | 2 | 3 |
| S5: m = y; | 1 | 2 | 2 |
| S6: else if (x<z) | 2 | 2 | 3 |
| S7: m = y;  //bug | 1 | 2 | 2 |
| S8: else | 3 | 0 | 0 |
| S9: if (x>y) | 2 | 2 | 3 |
| S10: m = y; | 1 | 2 | 2 |
| S11: else if (x>z) | 2 | 2 | 3 |
| S12: m = x; | 1 | 2 | 2 |
| S13: print("Median:", m); | 3 | 1 | 7 |

As shown in Table 2, by analyzing the faulty program under test, static program fault localization features such as the number of lines of code, variable and operator counts, and branch path counts can be extracted.

**4.2 Construction of the Deep Learning Software Fault Localization Model**

*4.2.1 MLP-FL Model.* This paper proposes a fault localization model based on a multilayer perceptron, called MLP-FL (Multi-layer Perceptron Fault Localization). This model is designed to handle three different dimensions of fault localization features, allowing for complementary information between them. Due to its simplicity and ease of use, this model can effectively improve the time efficiency of fault localization models. The structure of the model is shown in Fig. 4:

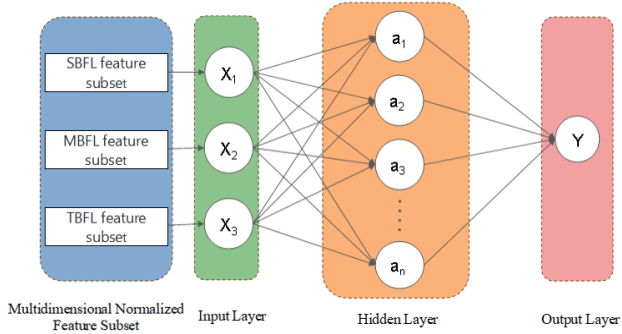

Fig. 4. Structure Diagram of the MLP-FL Model

As shown in Fig. 4, the structure of the MLP-FL model consists of an input layer, a hidden layer, and an output layer. The symbols $X_1$, $X_2$ and $X_3$ represent the three input neurons in the input layer, $a_1$ to $a_n$ represent the neurons in the hidden layer of the MLP-FL model, and Y represents the single neuron in the output layer. The output of the model is the fault prediction result, calculated using the following formula:

$$h_i = Sigmoid(W_l^i * X_l^i + b_i) \tag{12}$$

In Equation 12, $h_i$ represents the output of the *i*-th hidden layer node connected to the input layer neurons, $W_l^i$ represents the weight between the *i*-th input neuron and the *j*-th hidden layer neuron, $X_l^i$ represents the input value of the *i*-th input neuron, and $b_i$ represents the bias of the *j*-th hidden layer neuron. After calculating the outputs of all nodes in the hidden layer, the complete hidden layer can be represented as:

$$H = [h_1, h_2, \ldots h_n] \tag{13}$$

In Equation 13, H represents the complete hidden layer, and $h_i$ is the output of the i-th hidden layer neuron. The final output can be obtained using the following formula:

$$Y = \sigma(W_i * H + b) \tag{14}$$

In Equation 14, $Y$ represents the final output of the output layer neuron. Since the fault localization problem is essentially a binary classification problem, the output layer contains only one neuron, representing the fault localization result derived from the neural network model. The activation function used is denoted by $\sigma$, and in this study, the sigmoid function is selected as the activation function. $W_i$ represents the weight between the complete hidden layer and the output layer, g represents the complete hidden layer, and b represents the bias of the output layer.

$$Loss = -\frac{1}{N}\sum_{i=1}^{N} y_i \cdot \log(p(y_i)) + (1 - y_i) \cdot \log(1 - p(y_i)) \tag{15}$$

Since this paper addresses a binary classification problem for faulty statements, the model's loss function adopts the binary cross-entropy loss function, as shown in Equation 15, which represents the result value of binary cross-entropy used to evaluate the accuracy of the fault localization model's prediction results. For example, in actual testing, if the $i$-th statement in the dataset is a faulty statement, labeled as 1, if the model's predicted suspicion value approaches 1, the value of Loss should approach 0, indicating that this statement has been accurately predicted. Conversely, for statements without faults, if the predicted value $p(y_i)$ approaches 0, then the lower the value of the loss function Loss, the higher the accuracy of the model's prediction.

*4.2.2 RNN-FL Model.* Fig. 5 presents the framework structure of the RNN-FL (Recurrent Neural Network Fault Localization) model. As shown in the figure, the model input consists of three types of fault features: the SBFL (spectrum-based fault localization) feature subset, the MBFL (mutation-based fault localization) feature subset, and the TBFL (text-based fault localization) feature subset. In this study, mean imputation is used to normalize each feature group so that they all have the same length as the largest feature group. The normalized feature subsets are then input into the GRU for fault localization.

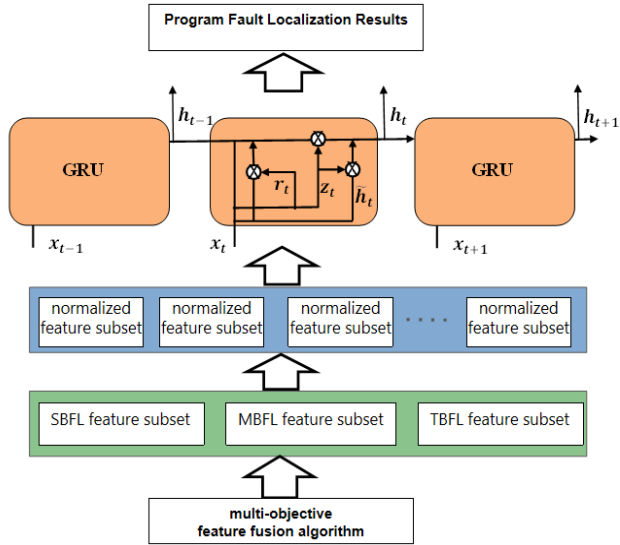

Fig. 5 Structure Diagram of the RNN-FL Model

The study employs a linear combination of multiple GRU units to adjust their internal states based on the features at each time step in the sequence, learning the differences between the feature groups. The feature inputs and past state information are concatenated as inputs for the gating mechanism and candidate hidden state, thereby achieving multi-dimensional fault localization. The complete multi-dimensional feature input $x_t$ can be represented as a feature vector of size n:

$$x_t = [f_1, f_2, f_3, \ldots f_n] \qquad (16)$$

where $f_n$ represents the n-th feature vector. The GRU unit calculates the new hidden state through the update gate, thereby dynamically adjusting and learning the fault feature sequence. The state calculation is as follows:

$$z_t = Sigmoid(W_z \cdot [h_{t-1}, x_t] + b_z) \qquad (17)$$

In Equation 17, $z_t$ represents the output of the update gate, which determines whether the past hidden state needs to be updated to the new hidden state. A value of 0 indicates the gate is completely closed, while a value of 1 indicates it is fully open. The sigmoid function can convert any real number to a value between 0 and 1, representing the on-off state of the gating unit. The state of each gate is determined by the current time step's feature input $x_t$ and the previous time step's hidden state $h_{t-1}$. The update gate performs a linear transformation of the input and hidden state under the influence of a weight matrix $W_z$ and bias vector $b_z$.

The reset gate output $r_t$ calculation process is shown in Equation 18:

$$r_t = Sigmoid(W_r x_t + U_r h_{t-1} + b_r) \tag{18}$$

where $W_r$ represents the weight matrix of the reset gate, and $U_r$ represents the weight matrix of the hidden state in the reset gate. By adding a bias term $b_r$, the model's flexibility can be adjusted to better adapt to multi-dimensional fault features. Subsequently, to better capture long-term dependencies in the time series and improve the accuracy of fault localization, a candidate hidden state vector $\tilde{h}_t$ needs to be calculated, as shown in Equation 19:

$$\tilde{h}_t = \tanh(W_h x_t + U_h(r_t \odot h_{t-1}) + b_h) \tag{19}$$

where $\odot$ represents element-wise multiplication, used to selectively retain or forget the past hidden state. $W_h$ represents the weight matrix of the candidate hidden state, and $U_h$ represents the weight matrix of the hidden state used to combine the hidden state of the previous time step. $b_h$ represents the bias term of the candidate state. The final state output of the GRU unit is given in Equation 20:

$$h_t = (1 - z_t) \odot h_{t-1} + z_t \odot \tilde{h}_t \tag{20}$$

where $h_t$ is the hidden state at the current time step, which is the output of the GRU unit. Depending on the output of the update gate, the GRU can selectively retain the past hidden state or update it to the new candidate hidden state, thereby obtaining the hidden state at the current time step.

The RNN-FL model uses the binary cross-entropy loss function to evaluate the accuracy of the fault localization model's predictions. To prevent overfitting, L2 regularization is applied in the binary cross-entropy loss function. Additionally, gradient descent is used to dynamically adjust the weight values, with the calculation process shown in Equation 21:

$$b_{new} = b_{old} - \alpha \cdot \nabla_\theta L \tag{21}$$

where $b_{new}$ represents the updated weight, $b_{old}$ represents the previous weight, $\nabla_\theta L$ is the gradient of the loss function with respect to the parameter. Due to the presence of gating mechanisms, the weight update involves adjusting multiple weight parameters for the update gate, reset gate, and hidden state by calculating the gradient of the loss function with respect to each parameter.

## 5 EXPERIMENT RESULT AND ANALYSIS

The primary objective of the experimental analysis conducted in this study is to address the following research questions (RQs):

RQ1: Compared to other state-of-the-art feature selection algorithms, does the multi-feature fusion algorithm achieve better performance and results than single-feature algorithms?

RQ2: How do fault localization models that employ the multi-objective feature fusion algorithm compare in accuracy and efficiency to models that do not use multi-objective algorithms?

RQ3: Do the MLP-FL and RNN-FL models offer advantages in improving fault localization performance compared to other advanced models, both deep learning-based and traditional?

RQ4: How does the improved deep learning fault localization model, based on the multi-objective feature fusion algorithm, perform on datasets beyond Defects4j, such as the PROMISE dataset?

## 5.1 Experimental Datasets

This study employs two widely accepted and readily available real fault databases: the Defects4j [25] and PROMISE datasets [26]. Defects4j is a database and extensible framework that provides real-world fault datasets to facilitate reproducible research in software testing. Defects4j contains six datasets: Chart, Closure, Lang, Math, Mockito, and Time. For the PROMISE dataset, data from six projects within the database were selected, including Java projects, C/C++ projects, C# projects, and others, covering areas such as game development, machine learning, artificial intelligence, and more. This dataset is widely used in fault localization research as a validation dataset.

The datasets were randomly divided into 10 parts using ten-fold cross-validation. In each iteration, one part is selected as the test set, while the remaining nine parts are used as the training set. The average result of the 10 fault localization runs is used as the final result.

## 5.2 Experimental Parameters and Evaluation Metrics

The parameter settings for the three multi-objective algorithms used in the experiments are listed in Tables 3 to 5, the parameter settings for the MLP-FL model can be found in Table 6, and the parameter settings for the RNN-FL model are provided in Table 7.

Table 3. NSGA-II Algorithm Parameters

| Description | Value |
| --- | --- |
| Population Size | 100 |
| Maximum Iterations | 200 |
| Crossover Probability | 0.6 |
| Mutation Probability | 0.1 |
| Crossover Distribution Index | 1 |
| Mutation Distribution Index | 1 |

Table 4. MOPSO Algorithm Parameters

| Description | Value |
| --- | --- |
| Population Size | 100 |
| Maximum Iterations | 100 |
| Maximum Archive Size | 100 |
| Velocity Update Parameters | 1.5 and 2 |
| Maximum Velocity | 1 |
| Minimum Velocity | -1 |

Table 5. MODE Algorithm Parameters

| Description | Value |
| --- | --- |
| Population Size | 100 |
| Maximum Iterations | 100 |
| Crossover Probability | 0.5 |
| Scaling Factor | 0.2 |

Table 6. MLP-FL Model Parameters

| Parameter Name | Value |
| --- | --- |
| Number of Iterations | 100 |
| Data Recording Interval | 10 |
| Number of Input Layer Neurons | 3 |
| Number of Hidden Layers | 1 |
| Number of Hidden Layer Neurons | 128 |
| Number of Output Layer Neurons | 1 |
| Optimization Algorithm | Adam |
| Learning Rate | 0.001 |

Table 7. RNN-FL Model Parameters

| Description | Value |
| --- | --- |
| Number of Iterations | 100 |
| Data Recording Interval | 10 |
| Number of Input Layer Neurons | 3 |
| Number of GRU Layers | 2 |
| Number of Hidden Layer Neurons | 64 |
| Number of Output Layer Neurons | 1 |
| Optimization Algorithm | Adam |
| Learning Rate | 0.001 |

## 5.3 Validating the Effectiveness of the Multi-dimensional Feature Method（RQ1）

To verify the effectiveness of the multi-dimensional feature selection method proposed in this paper, this section of the experiment utilizes the software fault dataset Defects4j. The verification metrics used are TOP-1, TOP-3, TOP-5, MAR (Mean Average Rank), and MFR (Mean First Rank). The verification subjects are four different feature selection schemes. For ease of representation, the method that selects features based solely on spectrum-based fault characteristics is abbreviated as SPE; the method that selects features based solely on mutation fault characteristics is abbreviated as MUT; the method that selects features based solely on text-based fault characteristics is abbreviated as TEX; and the multi-dimensional feature selection method that considers all three types of fault characteristics is abbreviated as MFS. These are respectively combined with two deep learning models, MLP and RNN, forming eight different feature selection model methods: MLP-SPE, MLP-MUT, MLP-TEX, MLP-MFS, RNN-SPE, RNN-MUT, RNN-TEX, RNN-MFS. Through their experimental performance on various performance metrics on the fault dataset Defects4j, the effectiveness of the multi-dimensional feature selection method proposed in this paper is verified. The best-performing feature extraction method within each deep learning model is highlighted in bold in the tables.

Table 8. Performance of Different Feature Extraction Methods on the Defects4j Dataset

| Feature Selection | Top-1 | Top-3 | Top-5 | MAR | MFR |
|---|---|---|---|---|---|
| MLP-SPE | 203 | 264 | 296 | 8.61 | 10.09 |
| MLP-MUT | 189 | 237 | 261 | 13.91 | 14.36 |
| MLP-TEX | 156 | 194 | 233 | 15.37 | 17.05 |
| **MLP-MFS** | **243** | **292** | **313** | **6.53** | **7.84** |
| RNN-SPE | 216 | 293 | 312 | 8.52 | 9.84 |
| RNN-MUT | 200 | 223 | 263 | 13.17 | 13.80 |
| RNN-TEX | 173 | 210 | 246 | 14.68 | 16.49 |
| **RNN- MFS** | **253** | **311** | **327** | **5.28** | **7.84** |

As shown in Table 8, the Top-1, Top-3, and Top-5 evaluation criteria represent the number of faults correctly predicted by the model. Fig. 6 and 7 display the performance of various feature selection algorithms in the MLP and RNN models, respectively.

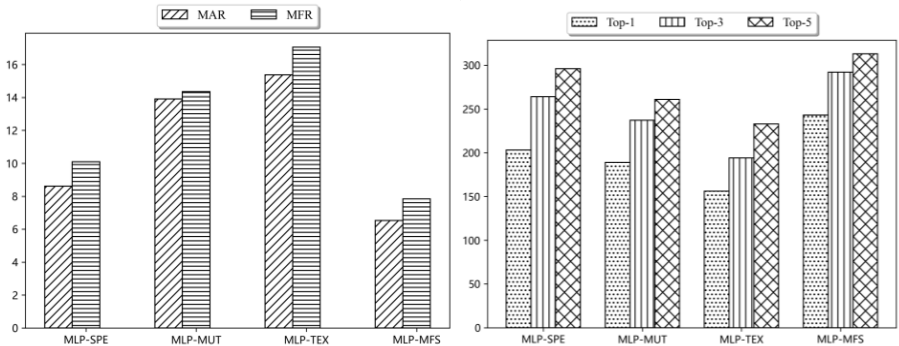

Fig. 6. Comparison of Fault Localization Performance for Different Feature Extraction Methods in MLP

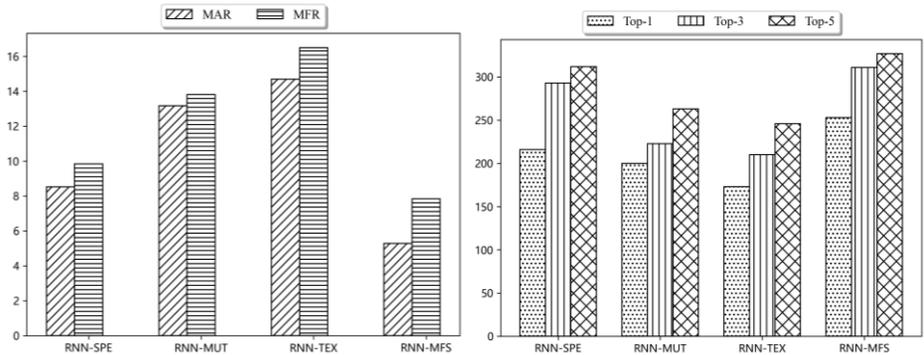
Fig. 7. Comparison of Fault Localization Performance for Different Feature Extraction Methods in RNN

From the results presented in the figures and tables, the following conclusions can be drawn:
1) In both MLP and RNN fault localization models, the proposed multi-dimensional feature extraction method outperforms SPE, MUT, and TEX in the Top-N metrics. The multi-dimensional feature extraction method also exhibits the best performance in the MAR and MFR metrics.
2) The MLP-MFS feature extraction method improves accuracy over the best single-feature method (MLP-SPE) by 19.7%, 10.6%, and 5.7% in the Top-1, Top-3, and Top-5 metrics, respectively. This indicates that the multi-dimensional feature extraction method performs better in higher-ranking positions. The performance of the four feature extraction methods in the RNN model is superior to that in the MLP model, suggesting that RNN models are more effective at processing program temporal information compared to MLP models.
3) Within the same model, the single-feature extraction methods ranked from best to worst are SPE > MUT > TEX, indicating that spectrum-based and mutation-based single features perform better. This also suggests that dynamic fault features contain more informative data, leading to better fault localization performance.

### 5.4 Validating the Effectiveness of the Multi-objective Feature Fusion Algorithm (RQ2)

To verify the effectiveness of the multi-objective optimization algorithms in multi-dimensional feature extraction, experiments were conducted on the Defects4j dataset. The NSGA-II (Nondominated Sorting Genetic Algorithm-II), MOPSO (Multi-objective Particle Swarm Optimization), and MODE (Multi-Objective Differential Evolution) algorithms were combined with the MLP-FL and RNN-FL fault localization models, and the fault localization time (in seconds) was measured across six data subsets.

Table 9. Time Performance (s) of Different Optimization Algorithms on the Defects4j Dataset

| Dataset | NSGA-II | | MOPSO | | MODE | | -- | |
|---|---|---|---|---|---|---|---|---|
| | MLP | RNN | MLP | RNN | MLP | RNN | MLP | RNN |
| Chart | 167 | 182 | **23** | 36 | 56 | 77 | 327 | 438 |
| Closure | 829 | 903 | **173** | 193 | 298 | 374 | 1247 | 1581 |
| Lang | 334 | 427 | **83** | 99 | 123 | 162 | 536 | 742 |
| Math | 386 | 461 | **105** | 132 | 175 | 193 | 604 | 670 |
| Mockito | 208 | 240 | **49** | 64 | 83 | 89 | 459 | 528 |
| Time | 90 | 136 | **22** | 31 | 44 | 49 | 126 | 199 |

Table 9 shows that on the Defects4J dataset, the MLP neural network outperforms the RNN in terms of time efficiency. The choice of multi-objective optimization algorithm significantly affects performance, with MOPSO performing best. Particularly on the complex data subset Closure,

NSGA-II took 3.79 times longer than MOPSO. Incorporating multi-objective optimization algorithms into deep learning fault localization models significantly improves time efficiency, confirming their adaptability and effectiveness. Fig. 8 shows the average time performance of different multi-objective optimization algorithms in MLP and RNN.

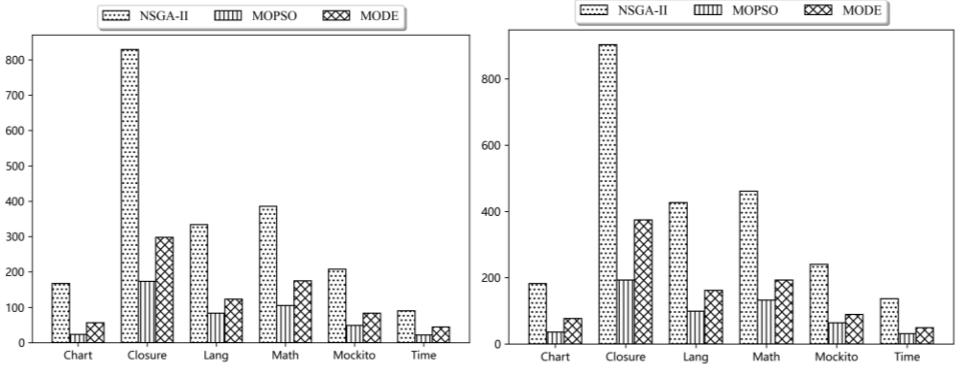

Fig. 8. Time efficiency of Different Multi-Objective Optimization Algorithms in MLP (left) and RNN (right)

From Fig. 8, it can be observed that the performance of NSGA-II and MODE on the dataset is significantly inferior to that of the MOPSO algorithm, with the performance gap widening on datasets with longer execution times. However, the performance difference between different neural network types is not significant. This indicates that, for the time efficiency of fault localization, the choice of multi-objective optimization algorithm has a greater impact than the choice of deep learning model. In practical applications, the appropriate multi-objective optimization algorithm should be selected based on specific circumstances.

Table 10 presents the average Area under the ROC Curve (AUC) values for multiple experiments using the NSGA-II, MOPSO, and MODE algorithms. The highest value within the same dataset is highlighted in bold, and the second-highest value is underlined.

Table 10. AUC Values of Different Optimization Algorithms on the Defects4j Dataset

| Dataset | NSGA-II | | MOPSO | | MODE | |
|---|---|---|---|---|---|---|
| | MLP | RNN | MLP | RNN | MLP | RNN |
| Chart | 0.914 | 0.921 | **0.928** | **0.928** | <u>0.925</u> | 0.924 |
| Closure | 0.907 | 0.919 | 0.921 | **0.938** | 0.907 | <u>0.931</u> |
| Lang | 0.951 | 0.944 | **0.968** | <u>0.965</u> | 0.953 | 0.948 |
| Math | 0.864 | 0.887 | 0.881 | **0.907** | 0.866 | <u>0.891</u> |
| Mockito | 0.941 | 0.914 | **0.946** | 0.940 | <u>0.943</u> | 0.929 |
| Time | 0.996 | 0.989 | <u>0.996</u> | 0.992 | **0.997** | 0.989 |

To visually represent the differences in AUC metrics between the methods, the data from Table 10 is displayed using a box plot in Fig. 9.

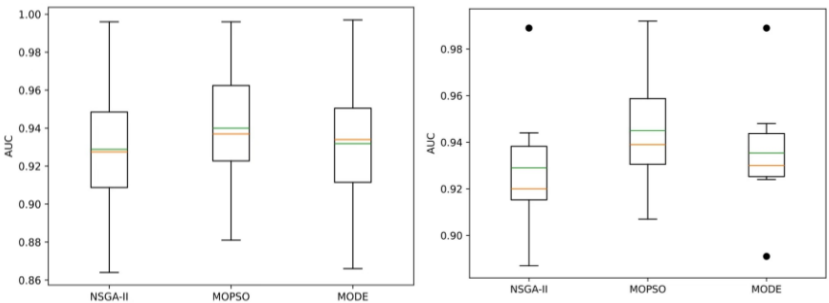

Fig. 9 AUC Performance of Different Optimization Algorithms in MLP (left) and RNN (right)

In Fig. 9, the green dashed line represents the mean, the yellow solid line represents the median, and the black dots represent outliers. The box plots show the AUC performance of the three multi-objective optimization algorithms in the two deep learning models. MOPSO performs best in five of the data subsets, with MODE performing best in one subset. It can also be seen that MOPSO is clearly superior to the other two multi-objective algorithms within the same deep learning model, except for the Time data subset where MODE has a slight advantage. This is likely due to the lower complexity of the Time data subset, which does not fully demonstrate the advantages of multi-objective optimization algorithms.

## 5.5 Validating the Effectiveness of the Deep Learning Fault Localization Models（RQ3）

To validate the effectiveness of the proposed method, two improved models introduced in this study, MLP-FL and RNN-FL, were compared with other software fault localization models, such as traditional models like Tarantula and Dstar, as well as more recent deep learning models like DeepFL. The effectiveness of each software fault localization model was evaluated.

Table 11. Overall Performance of Different Fault Localization Models on the Defects4j Dataset

| Model Name | Top-1 | Top-3 | Top-5 | MAR | MFR | AUC |
|---|---|---|---|---|---|---|
| Tarantula | 74 | 154 | 184 | 36.62 | 44.50 | 0.796 |
| Dstar | 83 | 158 | 194 | 32.91 | 39.36 | 0.831 |
| DeepFL | 195 | 260 | 288 | 7.61 | 9.09 | 0.880 |
| MLP-FL | 214 | 271 | 296 | 7.53 | 8.84 | 0.907 |
| **RNN-FL** | **221** | **276** | **303** | **6.58** | **8.22** | **0.928** |

From the analysis of the Table 11, the following observations can be made: The deep learning model methods outperform other software fault localization models in most experiments on the datasets. The RNN-FL model exhibits the best fault localization performance across all performance metrics, followed by the MLP-FL model, both of which surpass the traditional fault localization model Tarantula and the deep learning fault localization model DeepFL. Notably, in the Top-1 metric, MLP-FL identified 19 more faults than DeepFL, while RNN-FL identified 26 more faults than DeepFL.

The RNN-FL method shows the best performance across five datasets, particularly in the Lang and Chart subsets. Both methods perform better on the Defects4j dataset compared to other datasets, with DeepFL and MLP-FL consistently ranking second-best on the Defects4j dataset.

Table 12. Performance of Different Fault Localization Models on Different Subsets of the Defects4j Dataset

| Subset | Model Name | Top-1 | Top-3 | Top-5 | MAR | MFR |
|---|---|---|---|---|---|---|
| Chart | Tarantula | 6 | 14 | 15 | 8.94 | 9.66 |
| | Dstar | 8 | 14 | 15 | 7.63 | 9.01 |
| | DeepFL | 12 | 18 | 20 | 3.87 | 4.62 |
| | MLP-FL | 13 | 18 | 22 | 3.52 | 4.80 |
| | **RNN-FL** | **15** | **19** | **22** | **3.32** | **4.11** |
| Closure | Tarantula | 14 | 30 | 38 | 90.56 | 102.28 |
| | Dstar | 15 | 32 | 40 | 76.93 | 83.17 |
| Closure | DeepFL | 67 | 87 | 96 | 12.24 | 15.35 |
| | MLP-FL | 73 | 94 | 98 | 9.20 | 12.11 |
| | **RNN-FL** | **79** | **93** | **101** | **8.96** | **10.14** |
| Lang | Tarantula | 24 | 44 | 50 | 4.63 | 6.01 |
| | Dstar | 24 | 43 | 55 | 4.47 | 5.88 |
| | DeepFL | 46 | 53 | 59 | 2.24 | 2.98 |
| | MLP-FL | 51 | 54 | 59 | 2.10 | 2.72 |
| | **RNN-FL** | **52** | **57** | **60** | **1.96** | **2.53** |
| Math | Tarantula | 23 | 52 | 63 | 10.25 | 12.10 |
| | Dstar | 27 | 54 | 66 | 9.73 | 11.72 |
| | DeepFL | 58 | 83 | 91 | 4.22 | 5.34 |

|  | MLP-FL | 62 | 85 | 95 | 3.72 | 4.95 |
|  | **RNN-FL** | **63** | **85** | **95** | **3.53** | **4.68** |
|  | Tarantula | 7 | 14 | 18 | 30.22 | 25.64 |
|  | Dstar | 9 | 15 | 18 | 20.93 | 24.09 |
| Mockito | DeepFL | 12 | 19 | 22 | 13.92 | 16.78 |
|  | MLP-FL | 15 | 20 | 22 | 13.30 | 15.38 |
|  | **RNN-FL** | **12** | **22** | **25** | **11.75** | **13.66** |
|  | Tarantula | 6 | 11 | 13 | 15.96 | 18.87 |
|  | Dstar | 6 | 12 | 13 | 15.10 | 18.53 |
| Time | DeepFL | 13 | 16 | 17 | 13.92 | 16.62 |
|  | MLP-FL | 12 | 16 | 17 | 14.68 | 17.81 |
|  | **RNN-FL** | **13** | **17** | **17** | **11.92** | **14.56** |

Table 12 shows the fault localization results for various software fault localization models on the Defects4j dataset, using common fault localization standards such as Top-1, Top-3, Top-5, MAR, and MFR. The analysis indicates: Deep learning significantly outperforms traditional models in fault localization across all datasets. The three deep learning fault localization models exhibit strong performance on the Defects4j dataset.

Fig. 10 shows that the proposed MLP-FL and RNN-FL models perform well on the Defects4j dataset compared to the advanced DeepFL model.

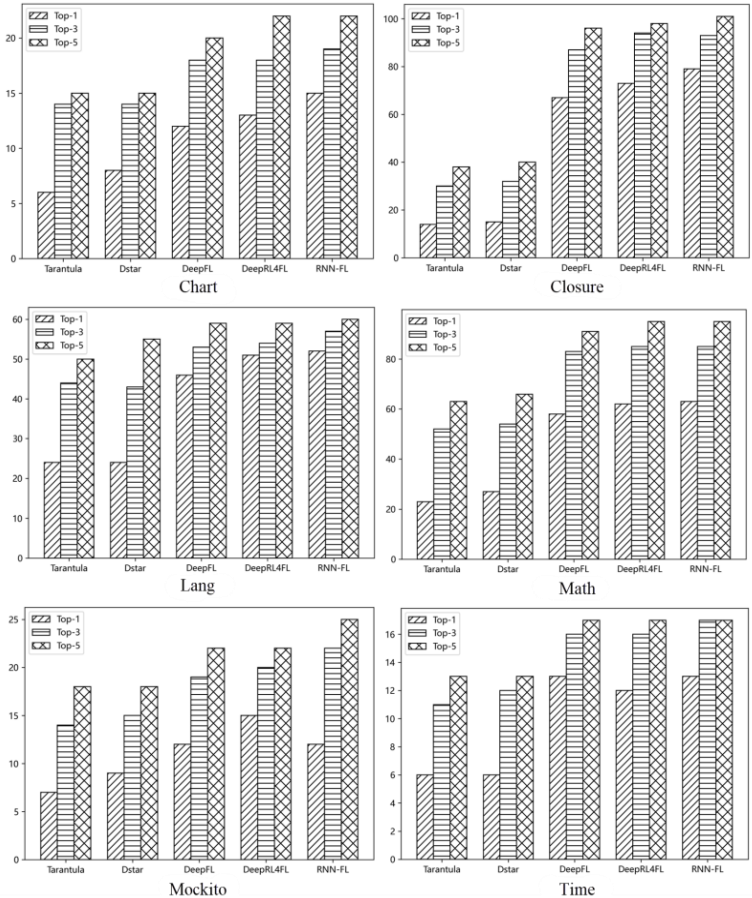

Fig. 10. Performance of Different Fault Localization Models on Various Data Subsets

By analyzing the experimental data and results, the proposed improved fault localization models are validated as effective, particularly the RNN-FL model, which shows a significant advantage. This suggests that deep learning models are suitable for fault localization problems and can effectively improve fault localization accuracy. The RNN-FL fault localization model, based on multi-objective feature fusion, is recommended for use in faulty programs.

## 5.6 Validating the Generalizability of the Deep Learning Fault Localization Models (RQ 4)

This section's experiments used six project fault datasets from the PROMISE fault library. In each experiment, one dataset was used as the training set, and the remaining datasets were used as test sets. The experiments were repeated 10 times, and the average was taken as the final result. The performance of the models was compared with Tarantula [27], Dstar [3], and DeepFL [7].

Table 13. Execution Results of the RNN-FL Model on the Ant-15 Faulty Program

| LOC | 20 | 40 | 60 | 80 | 100 | Rank |
|---|---|---|---|---|---|---|
| 1 | 4.02E-13 | 4.50E-13 | 4.32E-13 | 3.26E-13 | 2.87E-13 | 27 |
| 2 | 1.59E-14 | 3.97E-14 | 8.01E-14 | 5.84E-14 | 1.42E-13 | 19 |
| 3 | 5.19E-13 | 4.21E-13 | 1.47E-12 | 4.09E-13 | 3.28E-13 | 22 |
| 4 | 5.17E-15 | 1.79E-13 | 8.58E-12 | 2.01E-11 | 7.39E-12 | 8 |
| 5 | 4.05E-13 | 4.51E-13 | 1.44E-12 | 3.28E-13 | 2.88E-13 | 25 |
| 6 | 9.39E-15 | 3.17E-14 | 7.62E-14 | 7.26E-14 | 1.89E-13 | 17 |
| 7 | 5.4E-13 | 4.35E-13 | 1.52E-12 | 4.17E-13 | 3.36E-13 | 21 |
| 8 | 1.51E-14 | 1.20E-12 | 1.16E-11 | 9.93E-12 | 1.60E-12 | 10 |
| **9** | **3.56E-08** | **2.11E-05** | **0.000154** | **0.000745** | **0.000273** | **1** |
| 10 | 3.84E-14 | 1.20E-12 | 4.90E-12 | 2.89E-12 | 3.01E-13 | 14 |
| 11 | 4.02E-13 | 4.50E-13 | 1.43E-12 | 3.26E-13 | 2.87E-13 | 26 |
| 12 | 4.66E-15 | 4.34E-13 | 1.31E-11 | 1.57E-11 | 6.65E-12 | 6 |
| 13 | 1.48E-14 | 6.64E-13 | 8.42E-12 | 8.68E-12 | 1.32E-12 | 12 |
| 14 | 4.18E-15 | 2.40E-13 | 1.60E-11 | 8.25E-12 | 1.37E-11 | 2 |
| 15 | 3.84E-14 | 1.19E-12 | 4.76E-12 | 2.80E-12 | 2.93E-13 | 16 |
| 16 | 8.53E-14 | 1.14E-13 | 1.93E-13 | 1.12E-13 | 2.71E-13 | 20 |
| 17 | 1.03E-14 | 1.44E-12 | 2.60E-11 | 1.64E-11 | 2.51E-11 | 9 |
| 18 | 4.1E-13 | 4.51E-13 | 1.44E-12 | 3.28E-13 | 2.88E-13 | 24 |
| 19 | 4.68E-14 | 1.83E-12 | 1.67E-10 | 1.26E-10 | 1.17E-10 | 3 |
| 20 | 3.36E-14 | 9.68E-13 | 4.73E-12 | 2.79E-12 | 3.06E-13 | 13 |
| 21 | 3.84E-14 | 1.19E-12 | 4.76E-12 | 2.80E-12 | 2.93E-13 | 15 |
| 22 | 4.35E-13 | 4.56E-13 | 1.49E-12 | 3.49E-13 | 3.02E-13 | 23 |
| 23 | 1.22E-14 | 6.72E-13 | 1.06E-11 | 7.85E-12 | 1.39E-12 | 11 |
| 24 | 6.92E-15 | 8.74E-13 | 2.11E-11 | 5.86E-11 | 1.95E-11 | 5 |
| 25 | 3.73E-15 | 3.13E-13 | 7.29E-12 | 5.10E-12 | 3.52E-12 | 4 |
| 26 | 1.06E-14 | 5.12E-14 | 1.10E-13 | 7.69E-14 | 1.69E-13 | 18 |
| 27 | 5.82E-15 | 2.54E-13 | 1.10E-11 | 1.36E-11 | 4.66E-12 | 7 |

Table 14. Performance of Different Fault Localization Models on the PROMISE Dataset

| Model Name | Top-1 | Top-3 | Top-5 | MAR | MFR | AUC |
|---|---|---|---|---|---|---|
| Tarantula | 78 | 160 | 192 | 35.13 | 43.64 | 0.690 |
| Dstar | 74 | 165 | 203 | 33.37 | 38.97 | 0.722 |
| DeepFL | 183 | 223 | 259 | 10.25 | 11.82 | 0.879 |
| MLP-FL | **191** | 230 | 267 | 8.98 | 10.39 | 0.883 |
| RNN-FL | 188 | **237** | **271** | **7.69** | **9.44** | **0.891** |

Tables 13 and 14 show that in the cross-project software fault localization experiments, five models were compared: Tarantula, Dstar, DeepFL, MLP-FL, and RNN-FL. The comparison showed that MLP-FL and RNN-FL demonstrated significant advantages across multiple evaluation metrics, indicating that these models can more reliably distinguish faulty programs from normal ones, thereby improving the accuracy of cross-project software fault localization.

Based on the experimental results, it can be concluded that MLP-FL and RNN-FL have distinct advantages in cross-project software fault localization. They not only excel in the accuracy of identifying the first few faults but also quickly and accurately locate other faults, and they are more reliable in distinguishing between faulty and normal instances.

### 5.7 Analysis of Result Validity

In this study, multiple models were used in the fault localization experiments, and techniques such as cross-validation were employed to reduce randomness and ensure the reliability of the experimental results. Strict preprocessing and feature engineering were applied to the experimental data to ensure the quality and validity of the input data. Additionally, appropriate evaluation metrics were selected to assess model performance, and comprehensive result analysis and discussion were conducted. These measures enhance the credibility and reliability of the research. To demonstrate the effectiveness of the model, multiple datasets were used for fault localization comparison experiments, including the widely used Defects4J fault dataset for validation. Cross-project software fault localization experiments were conducted on the PROMISE dataset to showcase the model's generalizability. The software fault localization datasets were preprocessed to ensure there were no errors or missing data. However, these software fault datasets still face challenges such as limited data volume and insufficient data quality.

## 6. CONCLUSION

This paper presents a multi-objective fusion algorithm for fault localization feature extraction, effectively integrating spectrum-based, mutation-based, and text-based fault features. This approach addresses the problem of insufficient feature information in deep learning models and reduces computational redundancy through multi-objective optimization. Experimental results indicate that the proposed MLP-FL and RNN-FL models outperform state-of-the-art methods in both accuracy and computational efficiency. Specifically, experiments on the Defects4j dataset demonstrate that these models enhance fault localization accuracy, with the RNN-FL model achieving an average accuracy increase of 7.67% over the state-of-the-art model and an improvement of over 10% in Top-1 accuracy.